\documentclass{moriond}

\usepackage{amsmath}
\usepackage{amsfonts}
\usepackage{amssymb}
\usepackage{bm}
\usepackage{hyperref}
\usepackage{subcaption}
\usepackage{mathrsfs}
\usepackage{graphicx}
\usepackage{diagbox}
\usepackage{empheq}
\usepackage{multirow}
\usepackage{float}
\usepackage{graphicx}
\usepackage{ulem}
\def\be{\begin{equation}}
\def\ee{\end{equation}}
\def\bea{\begin{eqnarray}}
\def\eea{\end{eqnarray}}

\def\dd{\mathrm{d}}
\def\dM{\mathrm{M}}
\def\dS{\mathrm{S}}
\def\dMbar{\overline{\mathrm{M}}}
\def\dSbar{\overline{\mathrm{S}}}
\def\nn{\nonumber}

\newcommand{\FP}{\mathop{\mathrm{FP}}_{B=0}}
\newcommand{\FPprop}{\mathop{\mathrm{FP}}_{B=0}\Box^{-1}_\text{ret}}

\newcommand\calO{{\mathcal{O}}}

\newcommand{\Photo}{}
\begin{document}
\vspace*{4cm}
\title{Gravitational-wave tails of memory at 4PN order}

\author{David Trestini$^{a,b,\ast}$ and Luc Blanchet$^{a}$}

\address{
$^{a)}$ 
${\mathcal{G}}{\mathbb{R}}\varepsilon{\mathbb{C}}{\mathcal{O}}$, Institut d'Astrophysique de Paris,\\
UMR 7095, CNRS, Sorbonne Universit{\'e}, 98\textsuperscript{bis} boulevard Arago, 75014 Paris, France.\\
$^{b)}$
Laboratoire Univers et Th\'eories, Observatoire de Paris,\\
Universit\'e PSL, Universit\'e Paris Cit\'e, CNRS, F-92190 Meudon, France.}

\maketitle\abstracts{We study a novel cubic nonlinear effect, the tails-of-memory, which consist of a combination of the tail effect (backscattering of linear gravitational waves against the curvature of spacetime generated by the source) and the memory effect (due to reradiation of gravitational waves by linear gravitational waves themselves). Our final result is consistent with a straightforward direct computation of the memory effect, but also involves  many non-trivial tail-like terms.
}
\section{Introduction}

The recent completion of the flux and phase evolution of gravitational waves generated by compact binary systems at fourth and a half post-Newtonian order (4.5PN),~\cite{4PNMoriond,phase4PN,flux4PN} necessitated the control of the nonlinear propagation effects of gravitational waves through spacetime from the source to the observer.
These effects are encoded in the relation between the radiative moments, which parametrize the multipolar structure of the waveform as measured by an asymptotic far-away observer, and the so-called canonical moments, which parametrize the \emph{linearized} vacuum metric outside the source. The canonical moments are in turn connected \emph{via} a known procedure to the parameters of the physical matter source. The relations between radiative and canonical moments generically contain instantaneous terms, that only depend on the canonical moments at retarded time $u=t-r/c$, but also \emph{hereditary} terms, which entail a dependence of the radiative moments on the whole past history of the source. 

Some well-known hereditary effects are: the tail terms, which arise at 1.5PN order beyond quadrupole radiation, due to an interaction between the ADM mass of the system $\dM$ and the canonical quadrupole $\dM_{ij}$; the memory effect at 2.5PN, due to the $\dM_{ij}\times \dM_{ij}$ interaction; and the tail-of-tail $\dM\times \dM\times \dM_{ij}$ at 3PN. In this contribution, we are interested in the $\dM\times\dM_{ij}\times\dM_{ij}$ interaction which enters at 4PN order, dubbed \emph{tails-of-memory}, which can be represented by the following three Feynman diagrams.
\begin{figure}[H]\centering
\begin{subfigure}[b]{0.25\textwidth}
\includegraphics[width=\linewidth]{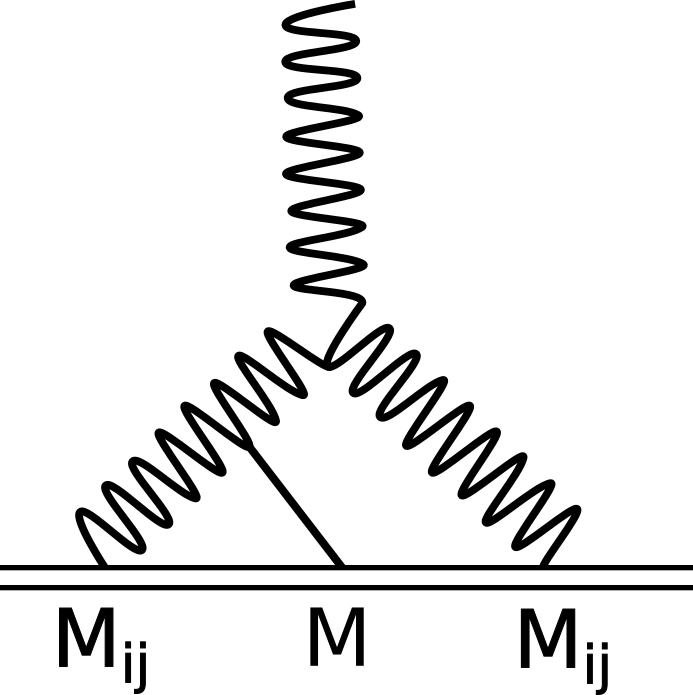}\caption{}\label{subfig:Feynman1}
\end{subfigure}
\qquad\qquad
\begin{subfigure}[b]{0.25\textwidth}
\includegraphics[width=\linewidth]{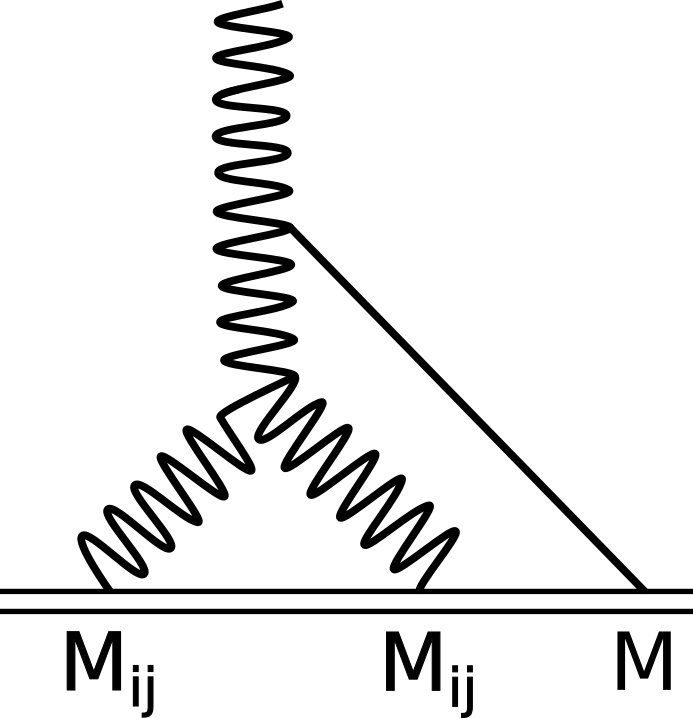}\caption{}\label{subfig:Feynman2}
\end{subfigure}
\qquad\qquad
\begin{subfigure}[b]{0.25\textwidth}
\includegraphics[width=\linewidth]{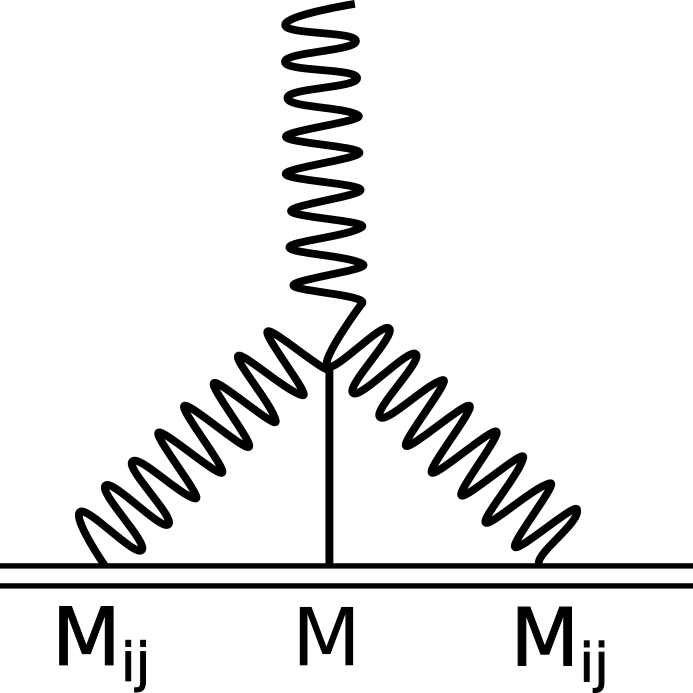}\caption{}\label{subfig:Feynman3}
\end{subfigure}
\caption{Feynman diagrams associated to the tails-of-memory.}
\label{fig:Feynman}
\end{figure}

\section{Multipolar post-Minkowkian setup}

In previous applications, the vacuum metric exterior to an isolated source was investigated using a multipolar post-Minkowskian (MPM) construction in \emph{harmonic} coordinates~\cite{BD86}, parametrized by two sets of canonical moments $\dM_L$ and $\dS_L$. The asymptotic behavior of the metric for $r \rightarrow \infty$ (and $u=$ const) then exhibited logarithms, which could ultimately be removed by a well-chosen coordinate transformation. Here, we present the \emph{radiative} MPM construction introduced in Ref.~\cite{B87}, see Ref.~\cite{TLB23} for a detailed comparison. This construction has the great advantage of automatically avoiding the appearance of far-zone logarithms. The starting point is to recast the vacuum field equations of general relativity in the form of a wavelike equation by defining the ``gothic'' metric deviation, namely $h^{\mu\nu} \equiv \sqrt{-g} g^{\mu\nu} - \eta^{\mu\nu}$ 
with standard notation. In a general coordinate system, the equations read
\begin{equation}\label{eq:EinsteinEquations} 
	\Box h^{\mu\nu} - \partial H^{\mu\nu} = \Lambda^{\mu\nu}\bigl[h, \partial h, \partial^2 h\bigr]\,,\end{equation}
where $\Box$ is the flat d'Alembertian operator, the divergence of the metric is $H^{\mu} \equiv \partial_\nu h^{\mu\nu}$ and we introduce the useful shorthand notation 
$\partial \theta^{\mu\nu} \equiv 2\partial^{(\mu} \theta^{\nu)}-\eta^{\mu\nu}\partial_\rho \theta^\rho$ for any $\theta^\mu$. With the MPM construction the radiative metric is formally decomposed to any PM order as $h ^{\mu\nu} = \sum_{n=1}^{+\infty} G^n h_n^{\mu\nu}$, where each of the coefficients is a functional of two sets of STF multipole moments $\dMbar_L$ and $\dSbar_L$ (different from $\dM_L$ and $\dS_L$ in harmonic coordinates) which parametrize the linear approximation $h_1^{\mu\nu}$ defined as
\begin{equation}\label{eq:h1}
	h^{\mu\nu}_1 \equiv h_{\text{can}\,1}^{\mu\nu}[\dMbar_L, \dSbar_L] + \partial \xi_1^{\mu\nu}\,, 
\end{equation}
where the ``canonical'' linear metric reads explicitly~\footnote{We use the symmetric-trace-free (STF) formalism, with $L=i_1\cdots i_\ell$ a multi-index with $\ell$ spatial indices, and $\hat{n}_L=\text{STF}[n_{i_1}\cdots n_{i_\ell}]$ the usual STF harmonics of order $\ell$. Similarly we shall denote $\hat{\partial}_L=\text{STF}[\partial_{i_1}\cdots \partial_{i_\ell}]$.}~\cite{Th80,BD86}
\begin{subequations}\label{eq:h1can}
	\small
	\begin{align}
		h_{\text{can}\,1}^{00}[\dMbar_L, \dSbar_L] &= - \frac{4}{c^2}\sum_{\ell \geqslant 0}\frac{(-)^\ell}{\ell !}\partial_L\left[\frac{1}{r}\,\dMbar_L\left(t-\frac{r}{c}\right)\right]\,,\\
		h_{\text{can}\,1}^{0i} [\dMbar_L, \dSbar_L]&=\frac{4}{c^3}\sum_{\ell \geqslant 1}\frac{(-)^\ell}{\ell !}\left\lbrace\partial_{L-1}\left[\frac{1}{r}\,\dMbar_{iL-1}^{(1)}\left(t-\frac{r}{c}\right)\right]+ \frac{\ell}{\ell+1}\epsilon_{iab} \partial_{aL-1}\left[\frac{1}{r}\,\dSbar_{b L-1}\left(t-\frac{r}{c}\right)\right]\right\rbrace\,,\\
		h_{\text{can}\,1}^{ij}[\dMbar_L, \dSbar_L] &= - \frac{4}{c^4}\sum_{\ell \geqslant 2}\frac{(-)^\ell}{\ell !}\left\lbrace\partial_{L-2}\left[\frac{1}{r}\,\dMbar_{ijL-2}^{(2)}\left(t-\frac{r}{c}\right)\right]+ \frac{2\ell}{\ell+1}\partial_{aL-2}\left[\frac{1}{r}\,\epsilon_{ab(i}\dSbar_{j) b L-2 }^{(1)}\left(t-\frac{r}{c}\right)\right]\right\rbrace\,.
	\end{align}
\end{subequations}
It satisfies the harmonic gauge condition, $\partial_\nu h_{\text{can}\,1}^{\mu\nu}=0$. The gauge vector in~\eqref{eq:h1} is
\begin{equation}\label{eq:xi1}
	\xi_1^{\mu} = \frac{2\dM}{c^2} \eta^{0\mu} \ln\left(\frac{r}{b_0}\right)\,,
\end{equation}
with $b_0$ an arbitrary constant length scale. The role of this gauge transformation is to ensure that the retarded time $u=t-r/c$ is asymptotically a null coordinate when $r\to\infty$. 

Once the linearized solution is defined, the nonlinear corrections are readily obtained by injecting the PM expansion into the field equations and solving these equations iteratively. At any order $n$, the general equation to solve is
\begin{equation}\label{eq:EinsteinEquationOrderN} 
	\Box h_n^{\mu\nu} - \partial H^{\mu\nu}_{n} = \Lambda_n^{\mu\nu}\,,
\end{equation}
where $\Lambda_n^{\mu\nu}\equiv\Lambda_n^{\mu\nu}[h_1, \cdots, h_{n-1}]$ is built out from previous iterations. The construction of the solution is done in three steps. We first construct a particular retarded solution to the wave equation, satisfying $\Box u_n^{\mu\nu} = \Lambda_n^{\mu\nu}$, as
\begin{equation}\label{eq:unMPM}
	u_n^{\mu\nu} = \FPprop \biggl[\left(\frac{r}{r_0}\right)^{B} \Lambda_{n}^{\mu\nu}\biggr]\,,
\end{equation}
where	$\Box^{-1}_\text{ret}$ is the standard retarded integral operator, $r_0$ is an arbitrary length scale and taking the finite part ($\FP$) when the complex parameter $B\to 0$ takes care of the divergence of the multipole expansion when $r\to 0$.
We then construct $v_n^{\mu\nu}$ such that it satisfies $\Box v_n^{\mu\nu} = 0$ and $\partial_\nu v_n^{\mu\nu}=-\partial_\nu u_n^{\mu\nu}$.
Finally, one can prove~\cite{B87} that $\Lambda_n^{\mu\nu} = r^{-2}{k^\mu k^\nu} \sigma_n(u,\mathbf{n}) \,+\, \calO\left({r^{-3}}\right)$ when $r \to\infty$, $u=$ const, where $k^\mu = (1,\mathbf{n})$ is the outgoing Minkowskian vector. This allows us to define the gauge vector
\begin{equation}\label{eq:xinmu}
	\xi_n^{\mu\nu} = \FPprop \biggl[ \left(\frac{r}{r_0}\right)^{B} \frac{k^\mu}{2 r^2} \int_0^{+\infty} \dd \tau\, \sigma_n(u-\tau, \mathbf{n})\biggr]\,.
\end{equation}
The solution of the radiative MPM construction is then given by
\begin{equation}\label{eq:hnmunu}
	h_n^{\mu\nu} \equiv u_n^{\mu\nu}+v_n^{\mu\nu} +\partial \xi_n^{\mu\nu}\,,
\end{equation}
which satisfies (\ref{eq:EinsteinEquationOrderN}) by construction. The gauge transformation $\partial \xi_n^{\mu\nu}$ ensures that $u$ is a null coordinate to $n^\text{th}$ order and removes the far-zone logarithms at that order.

Note that the usual \emph{harmonic} MPM construction is exactly the same as the previous \emph{radiative} one, except that at each step $n\geqslant 1$, one sets $\xi_n^\mu = 0$, and our ``seed'' linearized metric~\eqref{eq:h1can} is parametrized by \emph{harmonic} canonical moments $\dM_L$ and $\dS_L$ (note that $\dM = \dMbar$ for the ADM mass). 
Finally one can prove that the metric in the \emph{radiative} MPM construction is free of far-zone logarithms at all post-Minkowskian orders~\cite{B87,TLB23}, such that the leading order contribution in $1/r$ in a transverse-traceless (TT) gauge has the multipolar structure
\begin{equation}\label{TTwf}
h_{ij}^\text{TT} = - \frac{4 G}{c^2 r} \perp_{ij,ab}^\text{TT}\sum_{\ell=2}^{+\infty} \frac{1}{c^\ell \ell !}\left( n_{L-2}\, \mathcal{U}_{abL-2}(u) - \frac{2 \ell}{c(\ell+1)} n_{cL-2} \,\epsilon_{cd(a} \,\mathcal{V}_{b)dL-2}(u)\right)+\mathcal{O}\left(\frac{1}{r^2}\right)\,,
\end{equation}
where $\perp_{ij,ab}^\text{TT}$ is the usual TT projection operator and $\mathcal{U}_L$ and $\mathcal{V}_L$ are the radiative moments. The radiative moments differ from the canonical moments by nonlinear terms,
\begin{equation}
	\mathcal{U}_L = \dM_L^{(\ell)} + \calO(G)\,,\qquad\mathcal{V}_L = \dS_L^{(\ell)} + \calO(G)\,,
\end{equation}
%

\section{The memory effect}

The ``memory'' in the waveform of any radiative source can be expressed as the following contributions to the \emph{mass-type} radiative moments $\mathcal{U}_L^\mathrm{mem}$ (but $\mathcal{V}_L^\mathrm{mem} = 0$), see \emph{e.g.}~\cite{Th92,BD92,Fav09}:
\begin{equation}\label{eq:ULmem}
\mathcal{U}_L^\mathrm{mem} = \frac{2 c^{\ell-2}(2\ell+1)!!}{(\ell+1)(\ell+2)} \int \dd \Omega  \, \hat{n}_L \int_{0}^{+\infty} \dd\tau\,\frac{\dd E_\mathrm{gw}}{\dd t\dd \Omega}(u-\tau) \,,
\end{equation}
where the effective power density of gravitational waves is given by 
\begin{equation}\label{dEgw}
\frac{\dd E_\mathrm{gw}}{\dd u\dd \Omega} = \lim_{r \rightarrow +\infty} \frac{r^2 c^3}{32\pi G} \,\dot{h}_{ij}^\text{TT} \,\dot{h}_{ij}^\text{TT} \,.
\end{equation}
For our purpose, we are interested in the dominant memory contribution due to the $\dM_{ij}\times\dM_{ij}$ interaction\,\footnote{Strictly speaking the memory is the non-oscillatory part (DC) of the waveform.} followed by the subdominant one $\dM\times\dM_{ij}\times\dM_{ij}$ arising at the 1.5PN order beyond leading order.
For this, we need the expression of the radiative quadrupole at 1.5PN order~\cite{BD92},
\begin{equation}\label{eq:UijLinAndTail}
\mathcal{U}_{ij} = \dM_{ij}^{(2)} + \frac{2G\dM}{c^3}\int_{0}^{+\infty} \dd \tau\, \dM_{ij}^{(4)}(u-\tau)  \left[\ln\left(\frac{c \tau}{2 b_0}\right) + \frac{11}{12}\right]+\mathcal{O}\left(\frac{1}{c^5}\right) \,.
\end{equation}
Injecting~\eqref{eq:UijLinAndTail} into the TT waveform~\eqref{TTwf} and then in~\eqref{dEgw}--\eqref{eq:ULmem}, and restricting attention to the $\dM_{ij}\times\dM_{ij}$ and $\dM\times\dM_{ij}\times\dM_{ij}$ interactions, a little tensor algebra leads to the result
%
%
%
\begin{align}\label{eq:UijmemMxMijxMijBeforeIBP}
	\mathcal{U}_{ij}^{\text{mem}} =& - \frac{2G}{7c^5} \int_0^{+\infty}\!\!\dd \tau\, \dM_{a\langle i}^{(3)}(u-\tau) \,\dM_{j\rangle a}^{(3)}(u-\tau)\nonumber\\ &- \frac{8G^2 \dM}{7c^8} \int_0^{+\infty}\!\!\dd \rho\, \dM_{a\langle i}^{(3)}(u-\rho) \int_0^{+\infty}\!\! \dd \tau \ln\left(\frac{c \tau}{2b_0}\right) \dM_{j\rangle a}^{(5)}(u-\rho-\tau) \,,
\end{align}
where we have neglected the term associated with the constant $\frac{11}{12}$ in~\eqref{eq:UijLinAndTail}, which is instantaneous, and does not contribute to the memory effect. Moreover, we can integrate the cubic term in~\eqref{eq:UijmemMxMijxMijBeforeIBP} by parts --- the surface term is actually a tail term, and can be discarded as it does not contribute to the memory effect --- and find
\begin{align}\label{eq:UijmemMxMijxMij}
	\mathcal{U}_{ij}^{\text{mem}} =& - \frac{2G}{7c^5} \int_0^{+\infty}\!\!\dd \tau\, \dM_{a\langle i}^{(3)}(u-\tau) \,\dM_{j\rangle a}^{(3)}(u-\tau)\nonumber\\ &+ \frac{8G^2 \dM}{7c^8} \int_0^{+\infty}\!\!\dd \rho\, \dM_{a\langle i}^{(4)}(u-\rho) \int_0^{+\infty}\!\! \dd \tau \ln\left(\frac{c \tau}{2b_0}\right) \dM_{j\rangle a}^{(4)}(u-\rho-\tau) \,.
\end{align}
%
%
%

\section{Obtaining the full cubic expression for the tails-of-memory}

In the radiative MPM iteration, the wave equation we need to solve can always be reduced
to a wave equation whose source term admits a definite multipolarity $\ell$, say
\begin{equation}\label{eq:generalWaveEquation}
	\Box \Psi_L = \hat{n}_L\,S(r,t-r/c)\,,
\end{equation}
where $S(r,u)$ is an arbitrary function of $r=\vert\mathbf{x}\vert$ and $u=t-r/c$, that verifies straightforward smoothness properties, and tends sufficiently rapidly to zero when $r\to 0$.\,\footnote{Namely $S(r,u) = \mathcal{O}(r^{\ell+5})$ when $r \rightarrow 0$ with $t-r/c$ kept fixed, see Theorem 6.1 of~\cite{BD86}. If this condition is not satisfied, we can multiply the source by the regularization factor $(r/r_0)^B$ with $B \in \mathbb{C}$, which ensures convergence when $\Re(B)$ is large enough, and take the finite part in the $B \rightarrow  0$ expansion in the end.} We define
\begin{equation}\label{eq:Ralpha}
	R_\alpha(\rho, s) \equiv  \rho^\ell \int_\alpha^\rho \dd\lambda\, \frac{(\rho-\lambda)^\ell}{\ell!}\left(\frac{2}{\lambda}\right)^{\ell-1} S(\lambda,s) \,,
\end{equation}
where $\alpha$ is an arbitrary constant. Then the solution of~\eqref{eq:generalWaveEquation} can be written as
\begin{equation}\label{eq:PsiLwithRalpha} \Psi_L = c \int_{-\infty}^{t-r}\dd s\, \hat{\partial}_L \left[ \frac{R_\alpha\left(\frac{t-r/c-s}{2}, s\right)-R_\alpha\left(\frac{t+r/c-s}{2}, s\right)}{r}\right]\,,
\end{equation}
see Eq.~(6.4) in~Ref.~\cite{BD86}. Thanks to this equation, and after an involved computation detailed in~\cite{TB23}, we obtain the radiative quadrupole $\mathcal{U}_{ij}$ in term of $\dMbar_{ij}$ and $\dM$. We then go back to the usual harmonic-type moments, that are necessary to relate our result to previous ones, using~\cite{TLB23}
\begin{align}\label{eq:momentRedefinition}
		\dMbar_{ij} =
		\dM_{ij}
		&\
		+ \frac{G\,\dM}{c^3}\,\dM_{ij}^{(1)}  \left[-\frac{26}{15}+2 \ln\left(\frac{r_0}{b_0}\right)\right]+ \frac{G^2\dM^2}{c^6}\,\dM_{ij}^{(2)} \left[\frac{124}{45}-\frac{52}{15} \ln\left(\frac{r_0}{b_0}\right) +2  \ln^2\left(\frac{r_0}{b_0}\right)  \right]\nonumber\\
		& \
		+ \frac{G^2 \dM}{c^8}
		\bigg[-\frac{8}{21}\,\dM^{}_{a\langle i}\dM_{j\rangle a}^{(4)}
		-\frac{8}{7}\,\dM^{(1)}_{a\langle i}\dM_{j\rangle a}^{(3)}
		-\frac{8}{9}\,\dM^{(3)}_{a\langle i}\dS_{j\rangle \vert a}^{}\bigg]+\mathcal{O}\left(\frac{1}{c^9}\right)\,.
	\end{align}
One must then take into account all the lower-order contributions to the radiative quadrupole in order to correctly implement this redefinition. Finally, we find that the tails-of-memory read
\begin{align}
\mathcal{U}_{ij}^{\dM\times\dM_{ij} \times \dM_{ij}} &= \frac{8 G^2\dM}{7 c^8}\Bigg\{\int_0^{+\infty} \!\dd\rho\,  \dM_{a \langle i}^{(4)}(u-\rho) \int_0^{+\infty} \!\dd \tau\,  \dM_{j \rangle a}^{(4)}(u-\rho-\tau)  \left[ \ln\left(\frac{c\tau}{2 r_0}\right) - \frac{1613}{270} \right]\nn\\
&\quad - \frac{5}{2} \int_0^{+\infty} \!\dd\tau \,  (\dM^{(3)}_{a \langle i} \dM^{(4)}_{j \rangle a})(u-\tau)\left[\ln\left(\frac{c\tau}{2 r_0}\right)+\frac{3}{2} \ln\left(\frac{c\tau}{2b_0}\right) \right]\nn\\
&\quad - 3  \int_0^{+\infty} \!\dd\tau\, (\dM^{(2)}_{a \langle i}  \dM^{(5)}_{j \rangle a})(u-\tau) \left[\ln\left(\frac{c\tau}{2r_0}\right)  +\frac{11}{12} \ln\left(\frac{c\tau}{2b_0}\right)  \right]\nn\\
&\quad-\frac{5}{2} \int_0^{+\infty} \!\dd\tau \, (\dM^{(1)}_{a \langle i}\dM^{(6)}_{j \rangle a})(u-\tau) \left[\ln\left(\frac{c\tau}{2r_0}\right)  + \frac{3}{10} \ln\left(\frac{c\tau}{2 b_0}\right) \right]\nn\\
&\quad - \int_0^{+\infty} \!\dd\tau\,(\dM_{a \langle i}\dM^{(7)}_{j \rangle a})(u-\tau) \left[\ln\left(\frac{c\tau}{2r_0}\right) - \frac{1}{4}\ln\left(\frac{c\tau}{2 b_0}\right)\right]\nn\\
&\quad - 2  \dM^{(2)}_{a \langle i} \int_0^{+\infty} \!\dd\tau\,  \dM^{(5)}_{j \rangle a}(u-\tau) \left[ \ln\left(\frac{c\tau}{2r_0}\right)+ \frac{27521}{5040} \right]\nn\\
&\quad- \frac{5}{2}\,  \dM^{(1)}_{a \langle i} \int_0^{+\infty} \!\dd\tau \, \dM^{(6)}_{j \rangle a}(u-\tau)  \left[\ln\left(\frac{c\tau}{2r_0}\right)+\frac{15511}{3150} \right]\nn\\
&\quad+ \frac{1}{2} \, \dM_{a \langle i} \int_0^{+\infty} \!\dd\tau \,\dM^{(7)}_{j \rangle a}(u-\tau)  \left[ \ln\left(\frac{c\tau}{2r_0}\right)  - \frac{6113}{756}\right]\,  \Bigg\}\,.
\end{align}
In particular, we recover in the first line the memory term of~\eqref{eq:UijmemMxMijxMij} but we also obtain  many non-trivial tail-like terms. This result was crucial for the completion of the 4PN waveform and 4.5PN flux for compact binary system on quasicircular orbits~\cite{phase4PN,flux4PN}. Finally note that the two constants $r_0$ and $b_0$ nicely vanish in the final results~\cite{phase4PN,flux4PN}. 
\section*{Acknowledgments}
The authors thank Laura Bernard, Guillaume Faye, Quentin Henry, Fran\c{c}ois Larrouturou and Stavros Mougiakakos for interesting discussions.

\end{document}